# Oxide 2D electron gases as a route for high carrier densities on (001) Si

Lior Kornblum,[1,2] Eric N. Jin,[1,2] Divine P. Kumah,[1,2] Alexis T. Ernst,[1,3] Christine C. Broadbridge,[1,3] Charles H. Ahn[1,2,4] and Fred J. Walker[1,2]

[1]Center for Research on Interface Structures and Phenomena, Yale University, New Haven, CT 06511, USA

[2]Dept. of Applied Physics, Yale University, New Haven, CT 06511, USA

[3]Department of Physics, Southern Connecticut State University, 501 Crescent Street, New Haven, CT 06515, USA

[4]Dept. of Mechanical Engineering & Materials Science, Yale University, New Haven, CT 06511, USA

Two dimensional electron gases (2DEGs) formed at the interfaces of oxide heterostructures draw considerable interest owing to their unique physics and potential applications. Growing such heterostructures on conventional semiconductors has the potential to integrate their functionality with semiconductor device technology. We demonstrate 2DEGs on a conventional semiconductor by growing $GdTiO_3$-$SrTiO_3$ on silicon. Structural analysis confirms the epitaxial growth of heterostructures with abrupt interfaces and a high degree of crystallinity. Transport measurements show the conduction to be an interface effect, with ~$9 \times 10^{13}$ cm$^{-2}$ electrons per interface. Good agreement is demonstrated between the electronic behavior of structures grown on Si and on an oxide substrate, validating the robustness of this approach to bridge between lab-scale samples to a scalable, technologically relevant materials system.

Oxide interfaces are attracting significant interest owing to their rich physics and technological potential.[1,2] One of the most intriguing aspects of these interfaces is the emergence of interface conduction attributed to the formation of a 2-dimensional electron gas (2DEG) at the interface between two insulating oxides.[3,4] All-titanate interfaces, between $SrTiO_3$ and rare-earth titanates ($RTiO_3$, RTO), have gained recent attention because of their ability to obtain high carrier densities.[5,6] Such interfaces offer a combination of a high carrier density with a relatively simple interface structure, owing to the continuity of the Ti-O sublattice across the interface. Out of several RTO-STO combinations which have been studied, such as $LaTiO_3$,[7,8] $NdTiO_3$[9] and $SmTiO_3$,[10,11] $GdTiO_3$-$SrTiO_3$ (GTO-STO) has drawn intense interest. This material system features a high density 2DEG,[5] ferromagnetism,[12] quantum oscillations,[13] and enables the electrostatic modulation of its high carrier concentration.[14,15]

Until recently, 2DEGs formed at oxide interfaces have been demonstrated with structures grown on oxide substrates, such as STO,[3] $(LaAlO_3)_{0.3}(Sr_2AlTaO_6)_{0.7}$ (LSAT),[5,12] and several others. Recently, we demonstrated the growth of oxide heterostructures forming a 2DEG on silicon,[16,17] using the $LaTiO_3$ (LTO)-STO materials system. Growing these functional oxides on silicon brings them closer toward future on-chip application in technological devices and increases their potential for integration with conventional circuits. The ability to grow these structures on silicon further provides a route for large scale manufacturing,[18] in contrast to oxide substrates that are typically <1 cm$^2$ in size. In addition, the thermal conductivity of Si is ~1.5 W·cm$^{-1}$·K$^{-1}$, compared to ~0.11 and ~0.05 W·cm$^{-1}$·K$^{-1}$ for the common oxide substrates STO and LSAT (respectively). This provides an advantage for Si over oxide substrates in terms of heat dissipation, a benefit for high power devices.

In this work, we present the growth of GTO-STO on Si using molecular beam epitaxy. This similar material system offers certain advantages over LTO-STO. The growth of LTO requires reducing the oxygen background pressure by ~×5 compared to the STO growth pressure; this was found to result in better structures and avoid the formation of the pyrochlore phase,[19,20] $La_2Ti_2O_7$. Here we demonstrate the growth of high quality GTO at the same oxygen pressure used for growing STO and the formation of 2DEGs at these heterostructures.

GTO-STO heterostructures are grown using a custom-built reactive molecular beam epitaxy (MBE) at a base pressure of ~5×10$^{-10}$ Torr. 99.99% pure Sr, Gd (Sigma Aldrich) and Ti (Alfa Aesar) are thermally evaporated using effusion cells in a molecular oxygen background of ~5×10$^{-7}$ Torr introduced by a leak valve and a (thermocouple) substrate temperature of 600°C. The growth is monitored using *in-situ* reflection high-energy electron diffraction (RHEED) operated at 10kV.

High-resistivity silicon wafers are used as substrates, in order to eliminate substrate contributions to lateral conduction. 2 inch (001) undoped float-zone Si wafers (>3000 Ω·cm, Virginia Semiconductor) are cleaned using standard procedures and then transferred to ultrahigh vacuum. The growth of the first 2.5 unit cells (uc) of STO on Si is a multistep process, which is described in detail elsewhere.[16, 21] Various thicknesses of GTO (x=0, 2, 5 and 10 uc) are grown on 4.5 uc STO-templated Si, and are capped with a 15 uc of a top STO layer. Films are also grown on ceramic 5×5 mm$^2$ (001) LSAT substrates (CrysTec GmbH) for comparison which underwent a 10 min cleaning at 800°C in an oxygen plasma prior to growth. The growth is done under similar conditions used in the growth on Si.

Structural analysis is performed using x-ray diffraction (XRD, Rigaku Smartlab) and a scanning-transmission electron microscope (STEM, FEI Osiris) operated at 200kV using a high-angle annular dark field (HAADF) detector. Cross-section TEM samples are prepared by conventional polishing and ion milling. Electrical transport is measured with a Physical Properties Measurement System (Quantum Design) using the van der Pauw geometry and magnetic fields of ±3 T. Contacting the heterostructures is done by sputtering Au on the corners of 5x5 mm$^2$ pieces that are scratched beforehand, in order to vertically contact all the layers in the heterostructures.

RHEED patterns acquired before and after the growth of the top 15uc STO over a 10uc GTO / 4.5uc STO / Si structure are presented in Figs. 1a and 1b (respectively). The GTO surface shows half order streaks,[22] attributed to the rotation of the oxygen octahedra,[23] similar to what is observed with LTO.[16] Both RHEED patterns show continuous, narrow streaks indicating a smooth, two-dimensional crystalline surface. STEM micrographs taken at different magnifications show continuous and crystalline layers (Fig. 2) with an abrupt interface with Si (Fig. 2b). Some interfacial mixing may be present at the oxide-oxide interfaces,[5] particularly at the interface of GTO with the topmost STO layer.

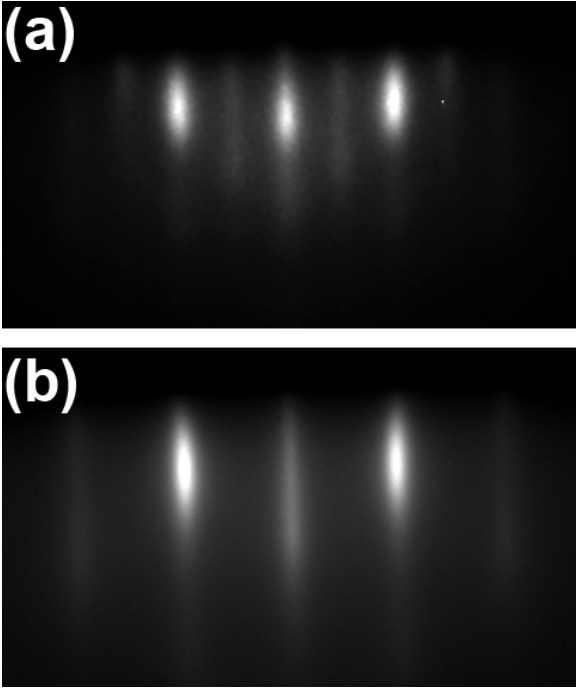

**Figure 1.** RHEED patterns from of a heterostructure comprised of 15/10/4.5 unit cells of STO-GTO-STO epitaxially grown on Si, acquired after (a) 10 uc of GTO and (b) top 15 uc STO, showing continuous streaks corresponding to a smooth crystalline surface.

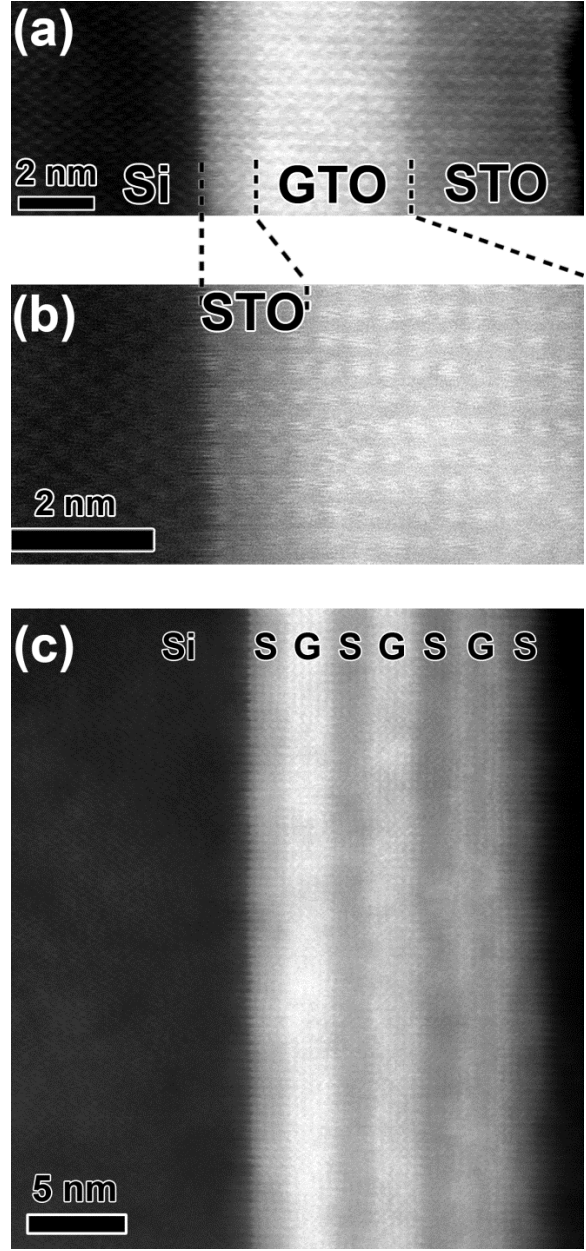

**Figure 2.** Cross section STEM micrographs of a heterostructure with 10 unit cells of GTO (Fig. 1) at (a) medium and (b) high magnification. (c) A STEM micrograph of a superlattice sample with 3 repetitions of (5uc GTO / 5uc STO) over 4.5uc of STO/Si. The symbols "S" and "G" denote STO and GTO, respectively. The intensity decrease from left to right is attributed to non-uniformities in the ion milling step of TEM sample preparation.

The (001) and (002) Bragg peaks of a 15/10/4.5 uc STO/GTO/STO structure grown on LSAT and on Si substrates (Fig. 3a,b) exhibit finite thickness oscillations, which further validate the abruptness of the interfaces. The curve was fitted using the GENX software,[24] yielding out of plane lattice parameters of 3.965±0.02Å and 3.93±0.02Å for GTO grown on LSAT and on Si, respectively. The errors represent fitting uncertainties. More accurate estimates of the lattice parameters are difficult to resolve due to the similarity of the lattice constants of STO and GTO. The fits are optimized with thicknesses of 5.69/4.12/1.76 nm and 5.87/3.93/1.76 nm [STO/GTO/STO(/substrate)] for LSAT and Si substrates (respectively), which are in good agreement with the nominal values of 5.86/3.92/1.76 nm. Asymmetric diffraction peaks [(103) and (113)] taken from the structure grown on Si result in an average in-plane lattice constant of 3.93±0.03 Å for the oxides, suggesting that the structure is relaxed from the Si in-plane lattice parameter and the GTO cell volume in this heterostructure is larger than for bulk GTO[25] by ~2.7%. Based on the analysis of the structural data, it is concluded that the same heterostructure grown on LSAT and on Si has a similar structure, with abrupt interfaces and a high degree of crystallinity.

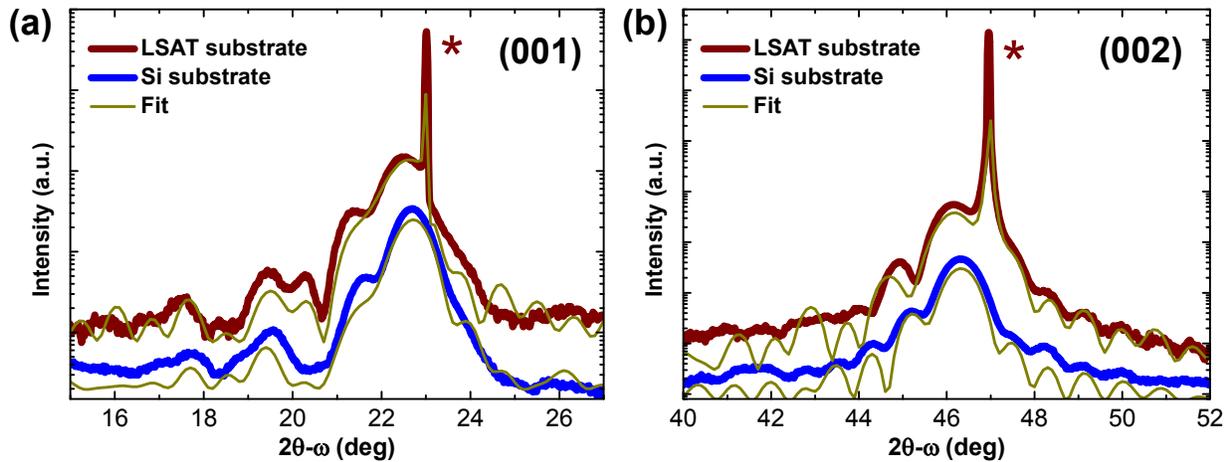

**Figure 3.** (color online) X-ray analysis of 15/10/4.5 unit cells of STO-GTO-STO grown on LSAT and on Si, showing (a) the (001) and (b) the (002) Bragg peaks. The curves are vertically shifted for clarity. The LSAT substrate peak is denoted by an asterisk.

The temperature dependent sheet resistances of STO-GTO-STO-Si structures with varying GTO thickness (Fig. 4a) show that the addition of a GTO layer inside an STO layer grown on Si (i.e. x>0) increases the sheet conductivity. This effect is independent of the GTO thickness, $x$, for 2 and 5 uc of GTO; a slightly more resistive structure is obtained for x=10 uc GTO. This indicates that the increased conductivity is a GTO-STO interface effect[5] and that bulk conduction in GTO is negligible in these structures. Similar transport properties are observed when an x=10 structure is grown on an LSAT substrate under the same conditions as on Si. The agreement between the structural (Fig. 3) and the electronic (Fig. 4) properties of heterostructures grown on LSAT and those grown on Si demonstrates the viability of this approach in bridging between lab-scale phenomena and a scalable materials system.

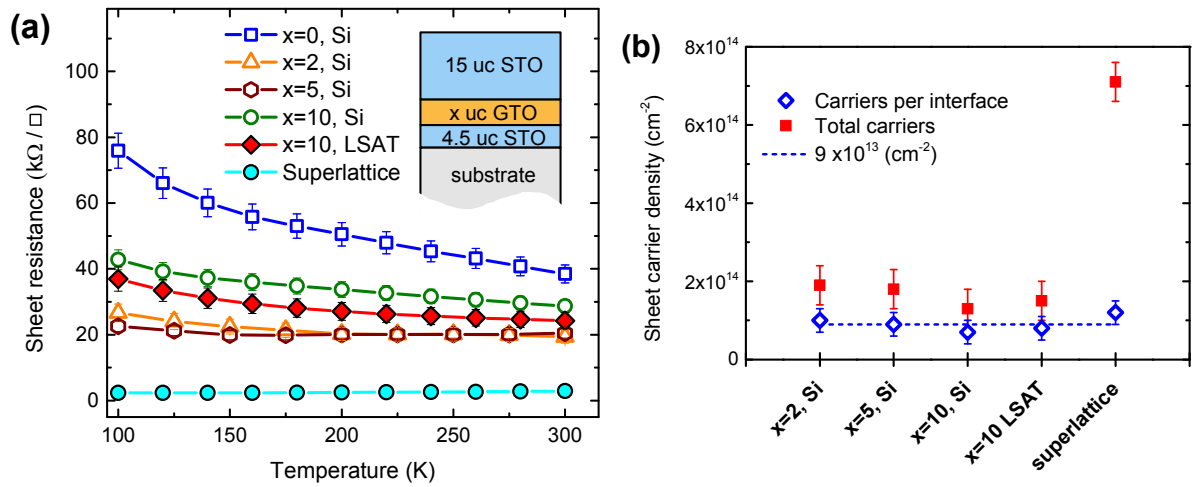

**Figure 4.** (color online) Transport properties of various configurations of STO-GTO-STO heterostructures. (a) Temperature dependence of the sheet resistance of the different stacks with x denoting the thickness of the GTO layer in unit cells, as depicted in the inset (error bars that are not visible are smaller than the symbols). (b) Sheet carrier density of the different stacks extracted from the linear Hall behavior at low temperatures. Closed squares describe the total sheet densities, and open diamonds describe the carriers per GTO-STO interface.

Although the insertion of GTO inside STO makes all samples more conductive, this increase is modest. It is therefore of interest to verify that the origin of the resistance reduction indeed stems from a GTO-STO interface effect. This effect was systematically established by Moetakef and co-workers,[5] but the difference in growth method and conditions warrants verification of the interfacial origin of the conduction. In order to check whether the resistance reduction is caused by a 2DEG formed at the GTO-STO interface, a superlattice sample was grown. The superlattice consists of the same total thickness of STO (19.5 uc), with three GTO layers (5 uc each) inside. A total of 15 uc of GTO were used in order to have an integer number of unit cells. This forms the structure [3×(STO/GTO)]/STO/Si (Fig. 2c) with all the layers being 5 uc thick, apart from the bottom STO, which is 4.5 uc thick. This structure manifests an increase in the total number of interfaces from 2 to 6, while keeping the STO thickness the same. The superlattice sample is considerably more conductive compared to the GTO-STO structures with just 2 interfaces (Fig. 4a), validating that the conduction is a GTO-STO interface effect.

Hall measurements of heterostructures grown on Si show a highly non-linear Hall resistance, $\rho_{xy}$, versus magnetic field behavior from room temperature down to ~100K, where it becomes linear. We interpret the non-linearity as a result of multiple-channel conduction in the heterostructures, consisting of the electron contribution of the 2DEG, and possible electron and holes contributions from the bulk of the Si and its interface with the bottom STO layer. While much lower in density compared to the 2DEG electrons, the carriers in the Si (~$10^9$ cm$^{-2}$) have mobilities higher by 2-3 orders of magnitude, resulting in the non-linear Hall behavior. Indeed, the non-linearity is not observed with structures grown on LSAT substrates, further validating the role of Si in the non-linearity of the Hall behavior.

Table I summarizes the sheet carrier densities of the samples, as extracted from the linear Hall data at low temperatures. The electron densities of the GTO-STO samples are ~$9\times10^{13}$ cm$^{-2}$ per GTO-STO interface. The superlattice structure shows a higher total sheet density of electrons, however, when scaled to the number of GTO-STO interfaces, agreement is obtained with the sheet density per interface observed in the other samples (Fig. 4b). For RTO-STO superlattices, the electron density scales with the number of interfaces regardless of STO thickness[10, 26] with an increase in mobility being observed in other work for 2DEGs grown on oxide substrates.[5] The carrier densities measured here are lower than those reported by Moetakef and co-workers,[5] $3\sim3.5\times10^{14}$ cm$^{-2}$. This disparity is attributed to structural imperfections and defects at the GTO-STO interface, caused by the lower growth temperature used here. The lower growth temperature of 600°C versus 950°C (ref. 12) is dictated by the need to preserve the STO-Si interface (Fig. 2b). GTO-STO interfacial defects may serve as localized traps for electrons and thus reduce the total number of carriers. Although the GTO-STO-Si structures exhibit somewhat lower carrier densities compared to similar structures on LSAT, we note that the densities reported here are comparable to high-quality LAO-STO grown on oxide substrates,[27-29] while encompassing the advantages of Si substrates.

**TABLE I**. Summary of the electronic properties of different GTO-STO heterostructures, extracted from their low temperature sheet resistance and Hall data.

|  | n (cm$^{-2}$) $\pm 0.5\times 10^{14}$ | # int | n/int (cm$^{-2}$) $\pm 3\times 10^{13}$ |
|---|---|---|---|
| x=2, Si | $1.9\times 10^{14}$ | 2 | $10\times 10^{13}$ |
| x=5, Si | $1.8\times 10^{14}$ | 2 | $9\times 10^{13}$ |
| x=10, Si | $1.3\times 10^{14}$ | 2 | $7\times 10^{13}$ |
| x=10 LSAT | $1.5\times 10^{14}$ | 2 | $8\times 10^{13}$ |
| Superlattice | $7.1\times 10^{14}$ | 6 | $12\times 10^{13}$ |

We note that using a higher (~×5) oxygen pressure during GTO growth compared to that used for LTO growth[16] results in a 10-fold reduction of the carrier density. We attribute this difference to the electronic contribution of oxygen vacancies localized at the LTO-STO interface.[28] In previous work, the contribution of bulk vacancies[30] was ruled out.[16] The formation of the interface vacancies may further be catalyzed by the rare-earth atoms during the initial sub-monolayer growth stages. Despite the extreme charge densities of the LTO-STO system, its sensitivity to oxygen may raise stability concerns during subsequent processing such as device fabrication.[31] In contrast, GTO-STO structures are shown here to be grown on Si without imposing further processing constraints, other than those required for STO-Si growth.

In summary, we demonstrate the epitaxial growth of GTO-STO heterostructures on Si and confirm their structure. Similar structure and electronic behavior are observed for GTO-STO grown on Si and LSAT substrates. It is shown that the 2D carrier density scales with the number of GTO-STO interfaces, thus confirming the origin of the conduction is a 2D electron gas. We conclude that GTO-STO heterostructures grown on Si show potential for the scalable integration of oxide 2DEGs with microelectronics technology.

**Acknowledgments**

This work was funded by the Office of Naval Research Multidisciplinary University Research Initiative (ONR-MURI) to support the EXtreme Electron DEvices (EXEDE) program, along with support from the National Science Foundation through NSF DMR-1309868 and NSF MRSEC DMR-1119826. Jesse Sabbagh and Kelly Woods are acknowledged for TEM sample preparation. The authors thank Marvin Wint and Timothy McHugh for valuable technical assistance.